\documentclass[12pt]{article}
\usepackage{graphicx}
\usepackage{tabularx}
\usepackage{epstopdf}
\usepackage{epsfig}
\usepackage{amsmath}
\usepackage{amsfonts}
\usepackage{amssymb}
\usepackage{colordvi}
\usepackage{slashed}
\usepackage{multirow}
\usepackage{subfigure}
\usepackage{rotating}
\usepackage{array}
\def\lsim{\mathrel{\raise.3ex\hbox{$<$\kern-.75em\lower 1ex\hbox{$\sim$}}}}
\def\gsim{\mathrel{\raise.3ex\hbox{$>$\kern-.75em\lower 1ex\hbox{$\sim$}}}}

\def\be{\begin{equation}}
\def\ee{\end{equation}}
\def\bea{\begin{eqnarray}}
\def\eea{\end{eqnarray}}

\begin{document}

\title{Light scalar dark matter at\\ neutrino oscillation experiments}

\author{Jiajun Liao,$^1$ Danny Marfatia,$^1$ and Kerry Whisnant$^2$\\
\\
\small\it $^1$Department of Physics and Astronomy, University of Hawaii-Manoa, Honolulu, HI 96822, USA\\
\small\it $^2$Department of Physics and Astronomy, Iowa State University, Ames, IA 50011, USA}
\date{}

\maketitle

\begin{abstract}

Couplings between light scalar dark matter (DM) and neutrinos induce a perturbation to the neutrino mass matrix. If the
DM oscillation period is smaller than ten minutes (or equivalently, the DM particle is heavier than $0.69\times10^{-17}$ eV), the fast-averaging over an oscillation cycle leads to a modification of the measured oscillation parameters. We present a specific $\mu-\tau$ symmetric model in which the measured value of $\theta_{13}$ is entirely generated by the DM interaction, and which reproduces the other measured oscillation parameters. For a scalar DM particle lighter than $10^{-15}$ eV, adiabatic solar neutrino propagation is maintained. A suppression of the sensitivity to CP violation at long-baseline neutrino experiments is predicted in this model.
We find that DUNE cannot exclude the DM scenario at more than
$3\sigma$ C.L. for bimaximal, tribimaximal and hexagonal mixing, while JUNO can rule it out at more than $6\sigma$ C.L. by precisely measuring both $\theta_{12}$ and $\theta_{13}$.
\end{abstract}

\newpage

\section{Introduction}

The existence of DM has been well established through
various cosmological and astrophysical observations. However, after decades
of experimental searches for DM, the particle
nature of DM is still unknown, and viable DM particle candidates span
an enormous mass range from fuzzy
DM~\cite{Hu:2000ke} to primordial black holes~\cite{Bird:2016dcv}.
Among the DM candidates, fuzzy DM with a mass range $1-10\times10^{-22}$~eV has
attracted much attention recently since it can resolve the small
scale crisis for standard cold DM due to its large de Broglie wavelength; see Ref.~\cite{Hui:2016ltb} and references therein. Constraints on fuzzy DM can be obtained 
from Lyman-$\alpha$ forest data and a lower limit of $20\times10^{-22}$~eV
at $2\sigma$ C.L. has been set from a combination of XQ-100 and
HIRES/MIKE data~\cite{Irsic:2017yje}, although a proper handling of the effect of quantum pressure and systematic uncertainties may relax the limit~\cite{Zhang:2017chj}.
Nevertheless, light scalar DM candidate of mass below a few
keV are generally expected in many extensions of the
Standard Model (SM). Examples include a QCD axion~\cite{axion}, moduli~\cite{moduli}, dilatons~\cite{dilaton}, and Higgs portal DM~\cite{Piazza:2010ye}.
Constraints on hot DM require that light scalar DM
cannot be produced thermally in the early universe. A popular production mechanism for generating
light scalar DM is the misalignment mechanism, in
which the fields take on some initial nonzero value in the early
universe, and as the Hubble expansion rate becomes comparable to the
light scalar mass, the DM field starts to oscillate as a coherent
state with a single macroscopic wavefunction~\cite{Suarez:2013iw}.

The properties of light scalar DM can be probed if they are coupled to
SM fermions, which induce a time variation to the masses of the SM
fermions due to the oscillation of the DM field.  Here we
consider the couplings between the light scalar DM and the SM neutrinos,
which were first studied in Ref.~\cite{Berlin:2016woy} by using
the nonobservation of periodicities in solar neutrino data. Constraints on light scalar DM couplings were also considered in Refs.~\cite{Krnjaic:2017zlz, Brdar:2017kbt} by using the data from various atmospheric, reactor and
accelerator neutrino experiments. In general, the
interactions between DM and neutrinos provide a small
perturbation to the neutrino mass matrix; a generic treatment of
small perturbations on the neutrino mass matrix is provided in
Refs.~\cite{Liao:2012xm, Liao:2015rma}. If the DM oscillation period is much smaller than the periodicity to which an experiment is sensitive, the oscillation probabilities get averaged, and a modification of the
oscillation parameters can be induced if the data are interpreted in the standard
three-neutrino framework. 

Evidence of time varying signals has been searched for in many neutrino oscillation experiments.  Super-Kamiokande finds no evidence for a seasonal variation in the atmospheric neutrino flux~\cite{Richard:2015aua}, and the annual modulation of the atmospheric neutrino flux observed at IceCube is correlated with the upper atmospheric temperature~\cite{ICseason}. 
Monthly-binned data from KamLAND indicate time variations in reactor powers~\cite{Fogli:2005qa}.
Tests of Lorentz symmetry via searches for sidereal variation in LSND~\cite{Auerbach:2005tq}, MINOS~\cite{Adamson:2008aa}, IceCube~\cite{Abbasi:2010kx}, MiniBooNE~\cite{AguilarArevalo:2011yi}, Double Chooz~\cite{Abe:2012gw}, and T2K~\cite{Abe:2017eot} data, are negative. Also, Super-Kamiokande~\cite{Yoo:2003rc} and SNO~\cite{Aharmim:2005iu} find no significant temporal variation in the solar neutrino flux with periods ranging from 
ten minutes to ten years. We therefore take the DM oscillation period to be smaller than ten minutes. 

In this work, we study the
modification of neutrino oscillation parameters due to light scalar DM--neutrino interactions. Since the predictions are flavor structure-dependent, we present a specific $\mu-\tau$ symmetric model in which the symmetry is
broken by light scalar DM interactions thus generating a nonzero
mixing angle $\theta_{13}$. We first examine the effects of this model on data from various neutrino experiments. We then study the potential to distinguish this
model from the standard three-neutrino oscillation scenario at the
future long-baseline accelerator experiment, DUNE, and the medium-baseline reactor experiment, JUNO.

The paper is organized as follows. In Section 2, we present a
model in which $\mu-\tau$ symmetry is broken by the DM
interactions. In Section 3, we examine the implications of this model for
the measured neutrino oscillation parameters. In Section 4, we simulate future neutrino
oscillation experiments to study the potential to distinguish this model from the standard scenario. We summarize our results in Section 5. In Appendix
A we calculate how the scalar DM interactions with neutrinos affect
neutrino mass and mixing parameters, and in Appendix B we determine how
the DM oscillations cause a shift in the effective neutrino oscillation parameters
measured in experiments.

\section{The Model}
The Lagrangian describing the interactions between light scalar DM and neutrinos can be written in the flavor basis as~\cite{Krnjaic:2017zlz, Brdar:2017kbt}
\bea
	\mathcal{L}=\bar{\nu}_{L\alpha} i\slashed\partial \nu_{L\alpha} -\frac{1}{2}m_0^{\alpha\beta}\overline{\nu_{L\alpha}^c}\nu_{L\beta}-\frac{1}{2}\lambda^{\alpha\beta}\phi\overline{\nu_{L\alpha}^c}\nu_{L\beta}+h.c.\,,
\eea
where $\alpha, \beta=e, \mu, \tau$, $m_0$ is the initial neutrino mass
matrix, and $\lambda$ is the coupling constant matrix. Since the light
scalar DM can be treated as a classical field, the nonrelativistic
solution to the classical equation of motion can be approximated as~\cite{Berlin:2016woy}
\bea
 \phi(x)\simeq\frac{\sqrt{2\rho_\phi(x)}}{m_\phi}\cos(m_\phi t-\vec{v}\cdotp\vec{x})\,,
 \eea
where $m_\phi$ is the mass of the scalar DM particle, $\rho_\phi\sim0.3 \text{ GeV}/\text{cm}^3$ is the local DM
density, and $v\sim10^{-3}$ is the virialized
DM velocity.  Since $v\ll1$, we neglect the spatial
variation in $\phi$ for neutrino oscillation experiments. 
In the presence of scalar DM interactions, the effective Hamiltonian for
neutrino oscillations can be written as
\begin{align}
H =&\frac{1}{2E_\nu}M^\dagger M+
\sqrt{2} G_F N_e \begin{pmatrix}
1 & 0 & 0
\\
0 & 0 & 0
\\
0 & 0 & 0
\end{pmatrix}\,,
\end{align}
where $N_e$ is the the number density of electrons. The effective mass matrix can be treated as the sum of an initial mass matrix and a small perturbation~\cite{Liao:2015rma}, i.e.,
\begin{align}
M=U_0^*\begin{pmatrix}
m_1^0 & 0 & 0
\\
0 & m_2^0 & 0
\\
0 & 0 & m_3^0
\end{pmatrix}U_0^\dagger + \mathcal{E}\cos(m_\phi t)\,,
\end{align}
where $U_0$ is the initial mixing matrix, $m_i^0$'s are the initial
neutrino eigenmasses, and the elements of the perturbation matrix are
\begin{align}
\mathcal{E}^{\alpha\beta}=\lambda^{\alpha\beta}\frac{\sqrt{2\rho_\phi}}{m_\phi}=0.0021 \text{ eV}\times\frac{\lambda^{\alpha\beta}}{10^{-22}}\times\frac{10^{-22} \text{ eV}}{m_\phi}\times\sqrt{\frac{2\rho_\phi}{0.3 \text{ GeV}/\text{cm}^3}}\,.
\end{align}
Note that the bounds in Fig. 1 of Ref.~\cite{Berlin:2016woy} only apply to a specific combination of $\lambda^{\alpha\beta}$ and $m_\phi$. Planck measurements yield $\sum m_\nu < 0.23$~eV at the 95\%~C.L.~\cite{Ade:2015xua}, which is much larger than the size of perturbation considered here.

We consider a model in which the initial mixing angle $\theta_{13}^0 = 0$ and the
measured $\theta_{13}$ value is generated by the DM interactions. 
In order to simplify our calculations, we specialize to models in which
the DM interactions only affect the masses at higher orders in the
perturbation, leaving them effectively unchanged. From the generalized
perturbation results of Ref.~\cite{Liao:2015rma}, 
we find that the most general
perturbation satisfying the latter requirement is of the form,
\begin{align}
\mathcal{E} = \begin{pmatrix}
0 & \sqrt2 \epsilon s_{23}^0 & \sqrt2 \epsilon c_{23}^0\\
\sqrt2 \epsilon s_{23}^0 & \epsilon^\prime \sin2\theta_{23}^0
& \epsilon^\prime \cos2\theta_{23}^0\\
\sqrt2 \epsilon c_{23}^0 & \epsilon^\prime \cos2\theta_{23}^0
& - \epsilon^\prime \sin2\theta_{23}^0
\end{pmatrix}\,.
\end{align}
As a further simplification, we assume the model is $\mu-\tau$
symmetric, i.e., $\theta_{23}^0 = \pi/4$. Then the perturbation becomes
\bea
\mathcal{E} = \begin{pmatrix}
	0 & \epsilon & \epsilon \\
	\epsilon & \epsilon^\prime & 0\\
	\epsilon & 0 & - \epsilon^\prime
\end{pmatrix}\,.
\eea\
With this perturbation, the shifts in all three angles are first order in
the small quantities $\epsilon$, $\epsilon^\prime$, and $\delta m_{21}$, where $\delta m_{ij}\equiv m_i-m_j$; since the eigenmasses are not shifted at leading order, we drop the superscript `{\scriptsize 0}' hereafter.
In Appendix~\ref{ap:second-order}, we show that the leading order corrections have amplitudes
\begin{align}
\delta\theta_{13}&\approx\frac{\sqrt{2}|\epsilon|}{\delta m_{31}}\,,
\label{eq:delta13}
\\
\delta\theta_{23}&\approx\frac{{\rm Re}(\epsilon^\prime)}{\delta m_{31}}\,,
\label{eq:delta23}
\\
\delta\theta_{12}&\approx
-\frac{{\rm Re}(\sqrt{2}\epsilon\epsilon^\prime\cos2\theta_{12}^0
	+(\epsilon^2-\epsilon^{\prime2}/2)\sin2\theta_{12}^0)}
{\delta m_{21}\delta m_{31}}\,,
\label{eq:delta12}
\\
\delta_{CP} &\approx {\rm arg}(\epsilon)\,.
\label{eq:deltaCP}
\end{align}
Note that $\delta\theta_{12}$ is second-order in the $\epsilon$'s and is 
therefore proportional to $\cos^2(m_\phi t)$, while
$\delta_{CP}$ depends only on the phase of $\epsilon$ and is constant, i.e.,
it is not affected by the DM oscillation. Both
$\delta\theta_{13}$ and $\delta\theta_{23}$ are dependent linearly on $\cos(m_\phi t)$.

\section{Effects on neutrino oscillation parameters}
In this section, we study how the neutrino
oscillation parameters are modified in our model, assuming the
period of the DM oscillation  ($\tau_\phi = 2\pi/m_\phi$) is short compared to the experimental resolution
of periodicity. 
Here we use the superscript `{\scriptsize 0}' to denote the initial oscillation parameters, and the superscript `{\scriptsize  $eff$}' to denote the effective parameters measured at neutrino oscillation experiments if the data are interpreted in the standard three-neutrino framework. For the parameters obtained after incorporating the DM perturbation, no superscript is used. 

\subsection{Short-baseline reactor experiments}
The leading oscillation probability for reactor antineutrinos (at a
Daya Bay-like distance) is $P = 1 - \sin^22\theta_{13} \sin^2\Delta_{31}$
where $\theta_{13} = \theta_{13}^0 + \delta\theta_{13}\cos(m_\phi t)$ and
$\Delta_{jk} = \delta m^2_{jk}L/4E$. Then expanding in powers of
$\delta\theta_{13}$ and averaging over a DM oscillation cycle, we get
\bea
\sin^22\theta_{13}\sin^2\Delta_{31} \simeq
\sin^2\Delta_{31}\left[\sin^22\theta_{13}^0
(1-4(\delta\theta_{13})^2)+2(\delta\theta_{13})^2\right]\,,
\eea
as found previously in Ref.~\cite{Krnjaic:2017zlz} . If $\theta_{13}^0=0$,
then $\sin^22\theta_{13}^{eff} = 2 (\delta\theta_{13})^2$, so
the angle being measured in these experiments is
\bea
\theta_{13}^{eff} \simeq \delta\theta_{13}/\sqrt2\,.
\label{eq:th13eff}
\eea

\subsection{Long-baseline appearance experiments}
\label{sc: lbl}
For long-baseline experiments, the formulas are more complicated. From Ref.~\cite{Barger:2001yr},
\bea
P(\nu_\mu\to\nu_e) = x^2f^2 + 2xyfg\cos(\Delta_{31}+\delta_{CP}) + y^2g^2\,,
\\
P(\bar\nu_\mu\to\bar\nu_e) = x^2\bar f^2 + 2xy\bar fg\cos(\Delta_{31}-\delta_{CP})
+ y^2g^2\,,
\eea
where $x = \sin\theta_{23}\sin2\theta_{13}$,
$y = \alpha\cos\theta_{23}\sin2\theta_{12}$, $\alpha = |\delta m^2_{21}/\delta m^2_{31}|$, 
$f, \bar f = \sin[(1\mp \hat A)\Delta_{31}]/(1\mp\hat A)$,
$g = \sin(\hat A\Delta_{31})/\hat A$,  $\hat A=|A/\delta m_{31}^2|$, and $A \equiv 2\sqrt2 G_F N_e E$.
Assuming $\theta_{13}^0 = 0$, and before averaging,
\bea
xf \approx 2f \delta\theta_{13} C \left[\sin\theta_{23}^0
+ C \cos\theta_{23}^0\delta\theta_{23}\right]\,,
\label{eq:xf}
\eea
where $C = \cos(m_\phi t)$. After averaging, the leading term for $x^2f^2$ is
\bea
x^2f^2 \approx 2 f^2 \sin^2\theta_{23}^0 (\delta\theta_{13})^2\,,
\eea
which is similar to the reactor case, i.e., the effective $\theta_{13}$
is $\delta\theta_{13}/\sqrt2$.

We can write $yg$ as
\bea
yg = y_0 g \frac{\cos\theta_{23}\sin2\theta_{12}}
{\cos\theta_{23}^0\sin2\theta_{12}^0}\,,
\label{eq:yg}
\eea
where $y_0 = \alpha\cos\theta_{23}^0\sin2\theta_{12}^0$. After explicitly putting in the perturbation, $yg$ becomes
\bea
yg \approx y_0 g \left[1 + 2C^2\delta\theta_{12}\cot2\theta_{12}^0
-C\tan\theta_{23}^0 \delta\theta_{23}\right]\,.
\eea
Combining Eqs.~(\ref{eq:xf}) and~(\ref{eq:yg}) and after averaging, the $xyfg$ term is
\bea
xyfg \approx
\left(\frac{\cos2\theta_{23}^0\delta\theta_{13}\delta\theta_{23}}
{\cos\theta_{23}^0}\right) y_0 f g \,,
\eea
where the term in parentheses replaces $x$ in the standard
expression. Note that this term is suppressed compared to the usual
case since it is proportional to two factors of the $\epsilon$'s
(assuming $\epsilon \sim \epsilon^\prime$), instead of just one --- the
term proportional to one factor of $\epsilon$ was linear in $C$ and
averaged to zero. For $\mu-\tau$ symmetry, $\theta_{23}^0 = \pi/4$
and the term vanishes completely. The upshot is that the effect
of the Dirac CP phase on $P(\nu_\mu\to\nu_e)$ and
$P(\bar\nu_\mu\to\bar\nu_e)$ is suppressed in long-baseline
neutrino oscillation appearance experiments.  Also, as shown in Appendix~\ref{ap:general},  this model predicts a suppression of the sensitivity to CP violation in all types of neutrino oscillation experiments.


\subsection{Medium-baseline reactor experiments}

For KamLAND and JUNO, the oscillation probability is
\bea
&P(\bar\nu_e \to \bar\nu_e) =
1 - \sin^22\theta_{13}\sin^2\Delta_{31}
- \cos^4\theta_{13}\sin^22\theta_{12}\sin^2\Delta_{21}
\nonumber\\
&\,\,\,\,\,\,\,\,\,\,\,\,\,\,\,\,\,\,\,\,\,\,\,\,\,\,\,+\sin^2\theta_{12}\sin^22\theta_{13}
\left[\frac{1}{2}\sin2\Delta_{21}\sin2\Delta_{31}+2\sin^2\Delta_{31}\sin^2\Delta_{21}
- \sin^2\Delta_{21}\right]\,.
\eea
For $\theta_{13}^0 = 0$, the angular factors after averaging over the DM
oscillations are
\bea
\langle \sin^22\theta_{13} \rangle &\approx&
2 (\delta\theta_{13})^2\,,
\label{eq:line1}
\\
\langle\sin^2\theta_{12} \sin^22\theta_{13} \rangle&\approx&
2 (\delta\theta_{13})^2 \sin^2\theta_{12}^0\,,
\label{eq:line2}
\\
\langle \cos^4\theta_{13} \sin^22\theta_{12} \rangle &\approx&
\sin^22\theta_{12}^0(1-(\delta\theta_{13})^2)
+ \sin4\theta_{12}^0\delta\theta_{12}\,.
\label{eq:factors}
\eea
Equations~(\ref{eq:line1}) and (\ref{eq:line2}) 
are identical to the standard case with $\sin^22\theta_{13}$ replaced by
$2(\delta\theta_{13})^2$, the same as for short-baseline reactor and
long-baseline accelerator experiments. In Eq.~(\ref{eq:factors}), the $\sin^22\theta_{12}^0$-dependent term on the right-hand side has a coefficient,
\bea
1 - (\delta\theta_{13})^2 = 1 - 2 \left(\delta\theta_{13}^{eff}\right)^2\simeq \cos^4\theta_{13}^{eff}\,.
\eea

Equation~(\ref{eq:factors}) also has an extra term proportional to $\delta\theta_{12}$. This
causes an effective shift in the measured value of $\theta_{12}$. To
determine how the shift depends on $\delta\theta_{12}$, neglect $\delta \theta_{13}$ and consider
\bea
\sin^22\theta_{12}^0 + 2\sin2\theta_{12}^0\cos2\theta_{12}^0\delta\theta_{12}
&=& \sin^22\theta_{12}^{eff}
\nonumber\\
&\approx& \sin^22\theta_{12}^0
+ 4\sin2\theta_{12}^0\cos2\theta_{12}^0\delta\theta_{12}^{eff}\,,
\eea
so that $\delta\theta_{12}^{eff} = \delta\theta_{12}/2$, and what one measures in this type of experiment is
\bea
\theta_{12}^{eff} = \theta_{12}^0 + \frac{1}{2}\delta\theta_{12}\,.
\label{eq:th12eff}
\eea

\subsection{Solar neutrinos}
In the SM scenario, solar neutrinos created in the center of the Sun undergo adiabatic evolution to the surface of the Sun, and travel to the Earth as an incoherent sum of the mass eigenstates. To not spoil the adiabatic evolution inside the Sun, we require that the period of the DM oscillation $\tau_\phi$ be much larger than the time in which neutrinos travel through the Sun, which is about $2.3$ seconds. This requirement restricts the mass of the scalar field: $m_\phi\ll1.8\times 10^{-15}$ eV. 

The three-neutrino survival probability for adiabatic propagation is
\bea
P(\nu_e \to \nu_e) &=&  \cos^2\theta_{13}\cos^2\bar{\theta}_{13}\left[\cos^2\theta_m\cos^2\bar{\theta}_{12}+\sin^2\theta_m^0\sin^2\bar{\theta}_{12}\right]+\sin^2\theta_{13}\sin^2\bar{\theta}_{13}
\nonumber\\
&=&
\frac{\cos^2\theta_{13}\cos^2\bar{\theta}_{13}}{2}\left[
1 + \cos2\theta_m \cos2\bar{\theta}_{12}\right]+\sin^2\theta_{13}\sin^2\bar{\theta}_{13}\,,
\eea
where
\bea
\cos2\theta_m = \frac{ \cos2\theta_{12}-\hat A_0 }
{\sqrt{( \cos2\theta_{12} - \hat A_0 )^2 + \sin^22\theta_{12}}}\,,
\label{eq:cos2theta}
\eea
with $\hat A_0= 2\sqrt2 G_F N_e^0 E/\delta m_{21}^2$, and $N_e^0$ is the electron number density at the point in the Sun where the
neutrino was created. Here $\theta_{ij}=\theta_{ij}^0 + \delta\theta_{ij}\cos(m_\phi t)$ and $\bar{\theta}_{ij}=\theta_{ij}^0 + \delta\theta_{ij}\cos(m_\phi t+m_\phi t_0)$ are the mixing angles at the production point in the Sun and at the Earth, respectively. They differ by a phase factor $m_\phi t_0$, where $t_0$ is the time traveled by neutrinos from the production point to the Earth.

Since $\theta_{13}^0 = 0$, expanding to the leading term,  we have
\bea
\cos^2\theta_{13} &\approx&
1-\delta \theta_{13}^2 \cos^2(m_\phi t)\,,
\nonumber\\
\sin^2\theta_{13}&\approx&
\delta\theta_{13}^2\cos^2(m_\phi t)\,,
\nonumber\\
\cos2\theta_{12} &\approx&
\cos2\theta_{12}^0-2\sin2\theta_{12}^0\delta \theta_{12}\cos^2(m_\phi t)\,,
\eea
and the corresponding barred quantities can be obtained by replacing the phase $m_\phi t$ with $m_\phi t+m_\phi t_0$. Also, to the leading order, we have 
\bea
\cos2\theta_m &\approx&
\cos2\theta_m^0+F\delta \theta_{12}\cos^2(m_\phi t)\,,
\eea
where $\cos2\theta_m^0$ has the same form as Eq.~(\ref{eq:cos2theta}), and 
\bea
F=\frac{  -2\sin2\theta_{12}^0}
{\sqrt{( \cos2\theta_{12}^0 - \hat A_0 )^2 + \sin^22\theta_{12}^0}}
-\frac{2\sin2\theta_{12}^0\hat A_0(\cos2\theta_{12}^0-\hat A_0)}{[( \cos2\theta_{12}^0 - \hat A_0 )^2 + \sin^22\theta_{12}^0]^{3/2}}\,.
\eea
%
If $P_0$ is the probability without the perturbation, i.e.,
\bea
P_0=\frac{1}{2}\left(1+\cos2\theta_m^0\cos2\theta_{12}^0\right)\,,
\eea
then keeping the leading correction for each $\delta \theta$, we have
\bea
P &\approx& P_0 +\frac{1}{2}\delta\theta_{12} [\cos 2\theta_{12}^0F\cos^2(m_\phi t)- 2\sin2\theta_{12}^0\cos2\theta_m^0\cos^2(m_\phi t+m_\phi t_0)] 
\\\nonumber
&-&\delta\theta_{13}^2 P_0  [\cos^2(m_\phi t)+\cos^2(m_\phi t+m_\phi t_0)] \,.
\eea
Because there is no interference term between $\cos^2(m_\phi t)$ and $\cos^2(m_\phi t+m_\phi t_0)$, we can average over them separately. Hence,
\bea
\langle P \rangle \approx P_0 + \frac{1}{2}\delta\theta_{12}\left(\frac{F}{2}\cos 2\theta_{12}^0 -\sin2\theta_{12}^0\cos2\theta_m^0\right) -\delta\theta_{13}^2 P_0\,.
\eea
By a similar calculation, the effective shifts in $\theta_{12}^{eff}$ and $\theta_{13}^{eff}$ lead to
\bea
P \approx P_0 + \frac{1}{2}\delta\theta_{12}^{eff} (F\cos 2\theta_{12}^0-2\sin2 \theta_{12}^0\cos2\theta_m^0)-2(\delta\theta_{13}^{eff})^2 P_0\,,
\eea
and we see that $\delta\theta_{12}^{eff} = \delta\theta_{12}/2$ and $\delta\theta_{13}^{eff} = \delta\theta_{13}/\sqrt{2}$, the same
as for medium-baseline reactor experiments. 

\subsection{Atmospheric neutrinos}
The survival probability of atmospheric neutrinos is 
\bea
P(\nu_\mu \to \nu_\mu) = 1 -
(\cos^4\theta_{13}\sin^22\theta_{23}
+ \sin^2\theta_{23}\sin^22\theta_{13})\sin^2\Delta_{31}\,.
\eea
For $\theta_{13}^0 = 0$, after averaging,
\bea
\langle \cos^4\theta_{13}\sin^22\theta_{23}\rangle &\approx&
\sin^22\theta_{23}^0(1 - (\delta\theta_{13})^2)
+2\cos4\theta_{23}^0(\delta\theta_{23})^2\,,
\\
\langle \sin^2\theta_{23}\sin^22\theta_{13} \rangle &\approx&
\sin^2\theta_{23}^0 2(\delta\theta_{13})^2\,.
\eea
Since the $(\delta\theta_{23})^2$ term is doubly suppressed, we have $\theta_{13}^{eff} \approx \delta\theta_{13}/\sqrt{2}$ and
$\delta\theta_{23}^{eff} \approx 0$. 
Thus,
\be
 \theta_{23}^{eff} \approx\theta_{23}^0\,.
\ee
This also applies to the long-baseline $\nu_\mu$
survival probability. Also, as shown in Appendix~\ref{ap:general},  matter effects do not change these results.

\section{Tests of the model in future neutrino experiments}
From the analytic analysis of the last section, we see that the constraints on this model mainly come from the measurement of $\theta_{13}^{eff}$ and $\theta_{12}^{eff}$. From Eqs.~(\ref{eq:delta13}) and~(\ref{eq:th13eff}), we have
\begin{align}
\theta_{13}^{eff} \simeq\frac{|\epsilon|}{\delta m_{31}}\,,
\label{eq:th13eff2}
\end{align}
and from Eqs.~(\ref{eq:delta12}) and~(\ref{eq:th12eff}), we have 
\begin{align}
\theta_{12}^{eff}\simeq\theta_{12}^0-\frac{{\rm Re}(\sqrt{2}\epsilon\epsilon^\prime\cos2\theta_{12}^0
	+(\epsilon^2-\epsilon^{\prime2}/2)\sin2\theta_{12}^0)}
{2\delta m_{21}\delta m_{31}}\,,
\label{eq:th12eff2}
\end{align}
where 
\begin{align}
\label{eq:dmnh}
\delta m_{31}=\sqrt{m_1^2+\delta m_{31}^{2}}-m_1\,, \\\nonumber
\delta m_{21}=\sqrt{m_1^2+\delta m_{21}^{2}}-m_1\,,
\end{align}
for the normal hierarchy, and 
\begin{align}
\delta m_{31}&=m_3-\sqrt{m_3^2-\delta m_{31}^{2}}\,, \\\nonumber
\delta m_{21}&=\sqrt{m_3^2-\delta m_{31}^{2}+\delta m_{21}^{2}}-\sqrt{m_3^2-\delta m_{31}^{2}}\,,
\end{align}
for the inverted hierarchy. 
Since the correction to $ \theta_{23}$ is doubly suppressed in the oscillation probabilities in this model, $\theta_{23}^{eff}$ remains maximal. 

We first study the sensitivity of long-baseline accelerator experiments to this model.
Since the currently running experiments, T2K and NO$\nu$A, have large experimental uncertainties, we consider the next-generation DUNE program. 
In our simulation, we use the GLoBES software~\cite{GLOBES} with the
same experimental configurations as in
Ref.~\cite{Liao:2016orc}. For the oscillation probabilities in the DM scenario, we modify the
probability engine in the GLoBES software by averaging the probabilities over a DM oscillation cycle numerically. We also use the Preliminary Reference Earth Model density profile~\cite{Dziewonski:1981xy}
with a 5\% uncertainty for the matter density. 

To obtain the sensitivities to the DM parameters at future long-baseline neutrino experiments, we simulate the data with the SM scenario in the normal hierarchy. Since the sensitivity to the Dirac CP phase is suppressed at such experiments, we conservatively choose $\delta_{CP}=0$. 
Also, due to the double suppression of the correction to $ \theta_{23}$, we choose $\theta_{23}=\frac{\pi}{4}$, which is within the $1\sigma$ range of the global fit~\cite{Esteban:2016qun}. We also adopt the other mixing angles and mass-squared differences from the best-fit values in the global fit, which are
\bea
&\sin^2\theta_{12}=0.307\,,\quad \sin^2\theta_{13}=0.022\,,\nonumber
\\
&\delta m_{21}^2=7.40\times 10^{-5} \text{ eV}^2\,, \quad \delta m_{31}^2=2.462\times 10^{-3} \text{ eV}^2\,.
\eea

We then test the DM model with the simulated data. We fix the initial mixing angles, $\theta_{23}^0=\frac{\pi}{4}$ and $\theta_{13}^0=0$. For $\theta_{12}^0$, we consider three benchmark values that are inspired by underlying discrete symmetries. Namely, $\theta_{12}^0=45^\circ$ for bimaximal (BM) mixing~\cite{BM}, $\theta_{12}^0=35.3^\circ$ for tri-bimaximal (TBM) mixing~\cite{TBM}, and $\theta_{12}^0=30^\circ$ for hexagonal (HG) mixing~\cite{HG}.
Since the masses are not affected at the leading order, we adopt the central values and uncertainties for the mass-squared differences from the global fit, i.e.,
\bea
\delta m_{21}^2=(7.40\pm 0.21)\times 10^{-5} \text{ eV}^2\,,
\quad
 |\delta m_{31}^2|=(2.462\pm 0.035)\times 10^{-3} \text{ eV}^2\,.
\eea
Also, since the long-baseline experiments are not sensitive to $\theta_{12}$, we impose a prior on $\theta_{12}^{eff}$ to account for constraints from the current global fit, i.e., $\sin^2\theta_{12}^{eff}=0.307\pm0.013$. We use Eq.~(\ref{eq:th12eff2}) to calculate the predicted value of $\theta_{12}^{eff}$. 
Then for a given $\theta_{12}^0$ and the lightest mass $m_1$ ($m_3$) for the normal (inverted) hierarchy, we scan over the magnitudes and phases of $\epsilon$ and $\epsilon^\prime$. We find that the phases of $\epsilon$ and $\epsilon^\prime$
only have a small effect on the $\chi^2$ value, which agrees with the analytical expectation that the measurement of the CP violation is suppressed.
We also marginalize over both the normal and inverted hierarchy for the tested DM scenario. We find that the $\chi^2$ value for the inverted mass hierarchy is always larger than that for the normal hierarchy for the same lightest mass. This is because the masses are not affected at the leading order and the mass hierarchy can be resolved with high confidence at DUNE~\cite{Acciarri:2015uup}.

\begin{figure}
	\centering
	\includegraphics[width=0.618\textwidth]{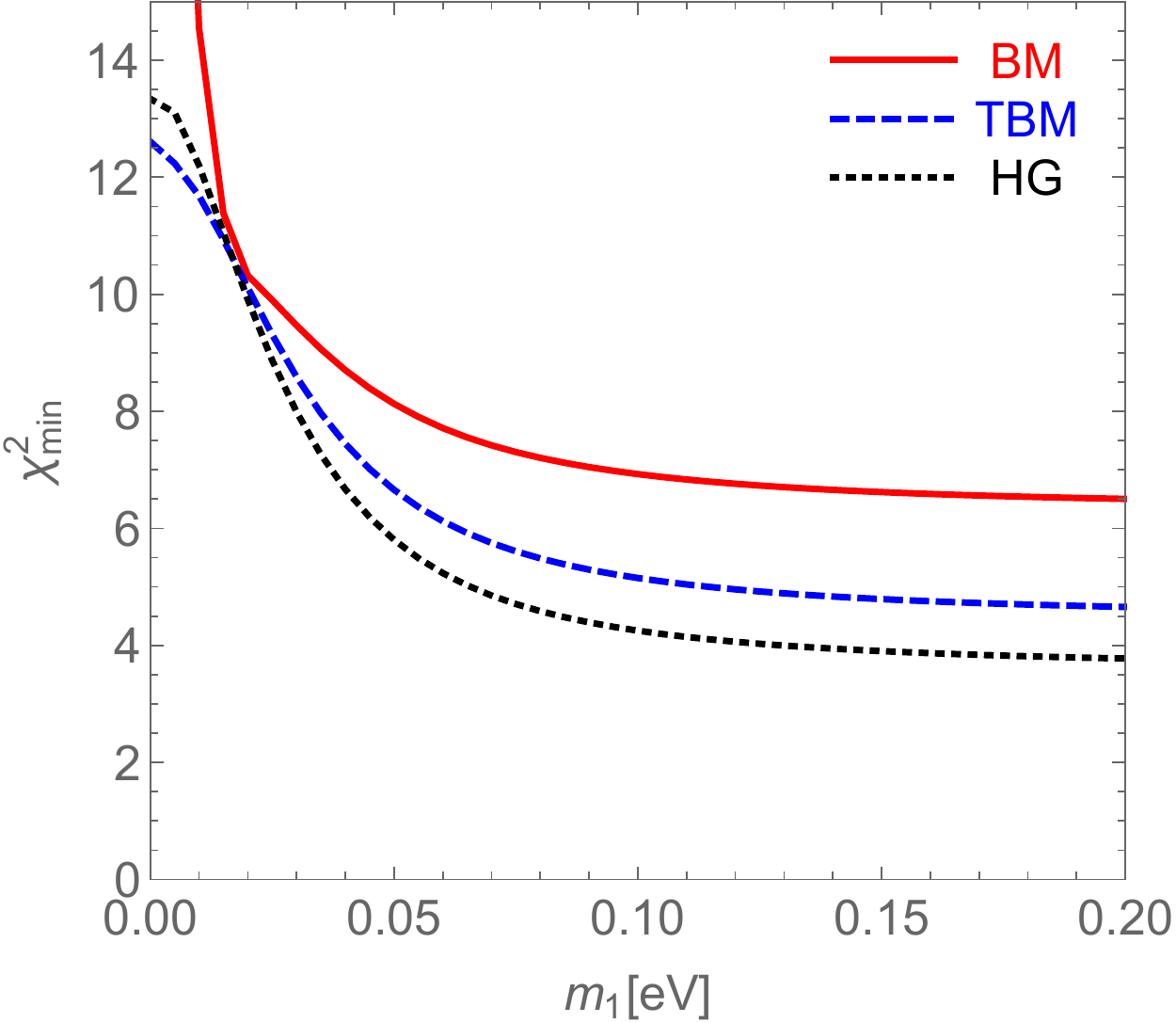}
	\caption{The minimum value of $\chi^2$ as a function of $m_1$ at DUNE. Here BM, TBM and HG correspond to $\theta_{12}^0=45^\circ$, $35.3^\circ$, and $30^\circ$, respectively.
	}
	\label{fig:dune-nh}
\end{figure}

\begin{figure}
	\centering
	\includegraphics[width=0.45\textwidth]{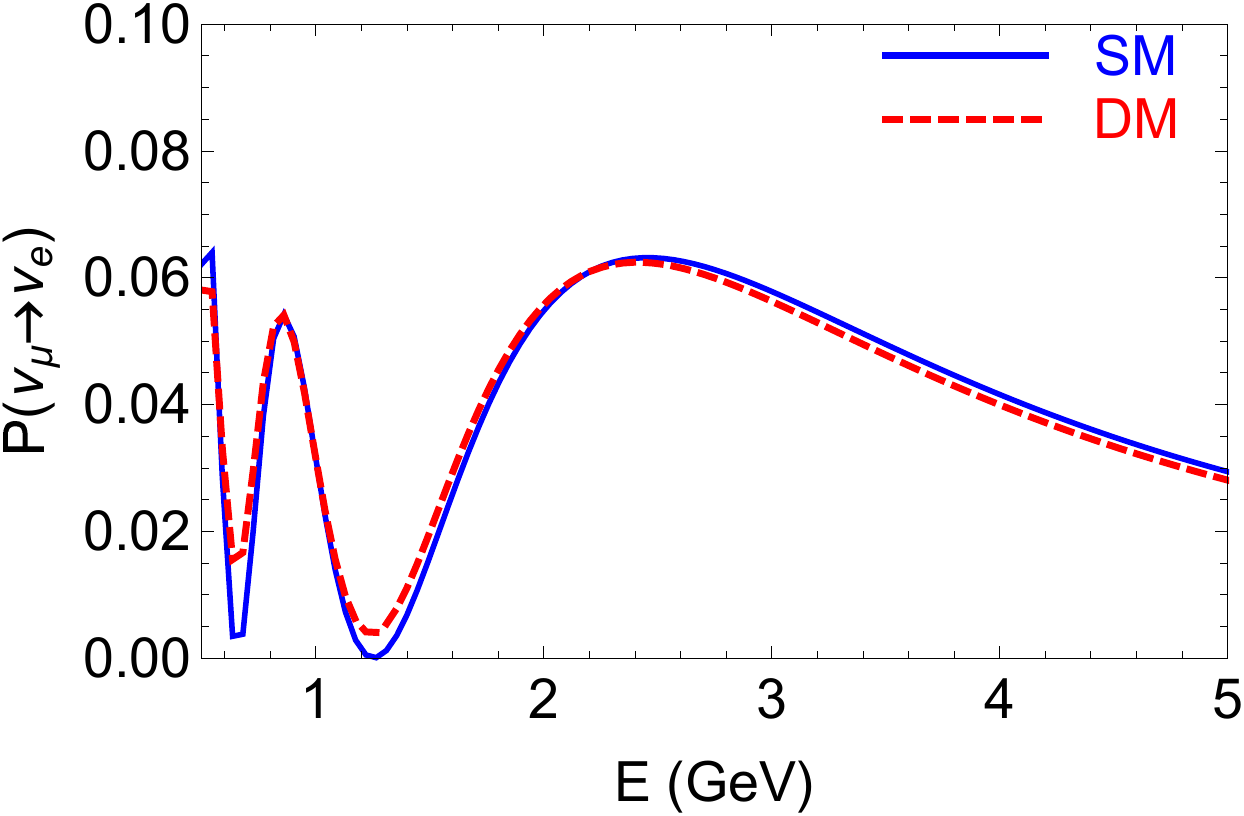}
    \includegraphics[width=0.45\textwidth]{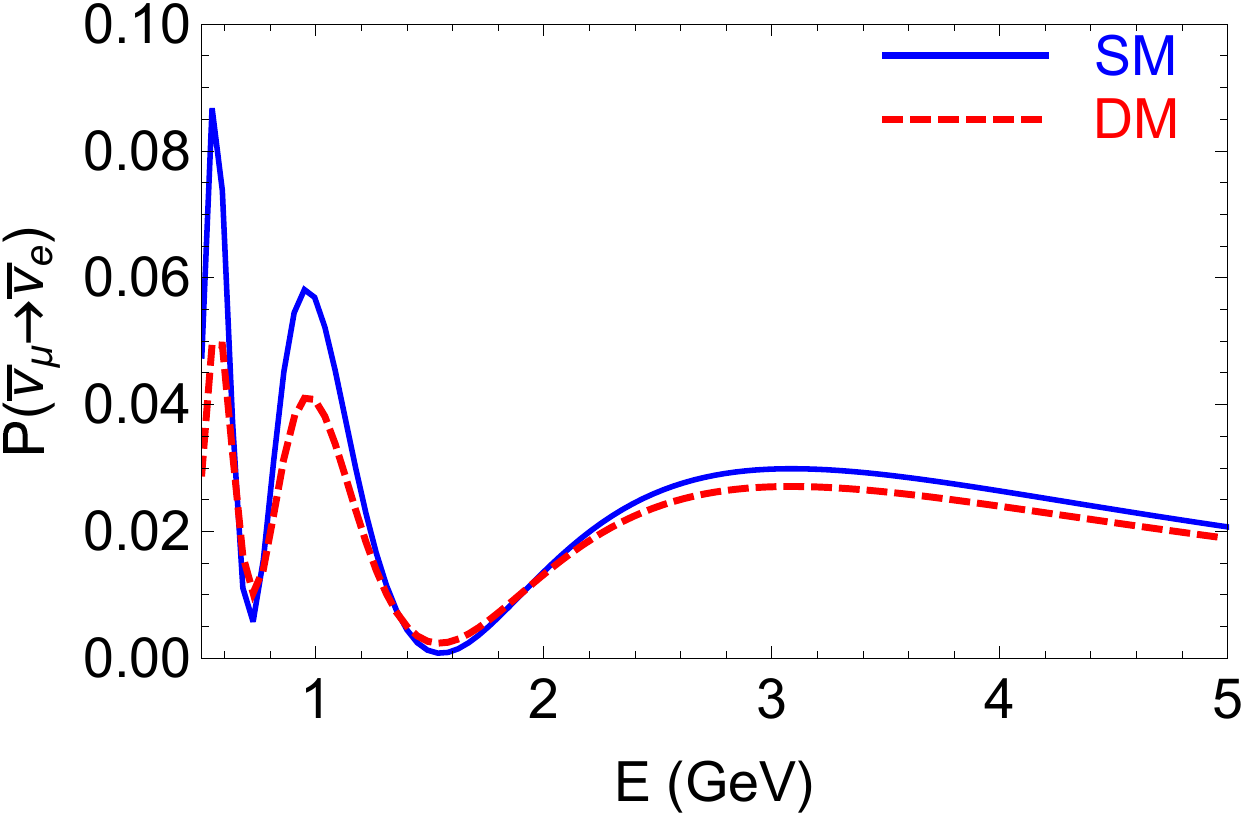}
	\caption{The appearance probabilities at DUNE. The oscillation parameters in the DM scenario are: $\theta_{12}^0=35.3^\circ$, $m_1=0.1$ eV, $\epsilon=0.0024\times e^{i0.7502}$ eV and $\epsilon'=0.00041\times e^{i1.278}$ eV. See the text for the other parameter values. 
	}
	\label{fg:dune-prob}
\end{figure}

The minimum value of $\chi^2$ as a function of $m_1$ is shown in Fig.~\ref{fig:dune-nh} for the three benchmark values of $\theta_{12}^0$.  As an illustrative example, we show the oscillation probabilities for $\theta_{12}^0=35.3^\circ$ and $m_1=0.1$ eV in the neutrino and antineutrino appearance channels in Fig.~\ref{fg:dune-prob}. We see that the DM oscillation curves overlap the SM curves sufficiently in both modes that a clear discrimination is not possible. From Fig.~\ref{fig:dune-nh} we see that DUNE alone cannot distinguish the DM scenario from the SM scenario at more than the $3\sigma$ C.L. if $m_1$ is greater than about 0.05 eV. We also see that as $m_1$ decreases, $\chi^2_\text{min}$ increases. This can be understood from Eqs.~(\ref{eq:th13eff2}) and~(\ref{eq:dmnh}). For a
smaller $m_1$, the magnitude of $\epsilon$ required to explain the measured $\theta_{13}$ becomes larger, and higher order corrections then break the degeneracies between the SM and DM scenarios.

Since future medium-baseline reactor experiments can make a high precision measurement of both $\theta_{12}$ and $\theta_{13}$, we study the sensitivity reach at JUNO. We use the GLoBES software to simulate the JUNO experiment. The experimental configuration is the same as that in Ref.~\cite{Liao:2017awz}, which reproduces the results of Ref.~\cite{An:2015jdp}. We use the same procedure for the long-baseline accelerator experiments except with no prior on $\theta_{12}^{eff}$, since JUNO can measure $\theta_{12}$  more precisely than the current experiments. For the lightest mass between 0 and 0.2~eV, we find that the minimum value of $\chi^2$ at JUNO is 47.6, 46.9 and 57.0,  with the initial mixing being BM, TBM and HG, respectively. Hence, JUNO can rule out this model with the three initial mixings at more than $6\sigma$ C.L.

\section{Summary}
We studied the effects of light scalar DM--neutrino interactions at various neutrino oscillation experiments. 
For a light scalar DM field oscillating as a coherent state, the coupling between DM and neutrinos  induces a small perturbation to the neutrino mass matrix. We consider the case in which the DM oscillation period is smaller than the experimental resolution of periodicity, i.e., ten minutes. After averaging the oscillation probabilities over a DM oscillation cycle, the perturbation to the neutrino mass matrix leads to a modification of the effective neutrino oscillation parameters if the experimental data are interpreted in the standard three-neutrino oscillation framework. 

Since the results depend on the flavor structure of the initial mass matrix and the perturbation matrix, we presented a specific $\mu-\tau$ symmetric model with DM interactions that do not affect the eigenmasses at the leading order. We examined the effects of this model on the effective oscillation parameters measured at various neutrino experiments. If the mass of the scalar field is lighter than $1.8\times 10^{-15}$~eV, then solar neutrinos propagate adiabatically. We find that all existing neutrino oscillation results can be explained in this model with a shift of the effective mixing angles --- the measured value of $\theta_{13}$ arises wholly from DM--neutrino interactions. The model also predicts a suppression of the CP violation at neutrino oscillation experiments.
 
We then studied the potential of DUNE and JUNO to
discriminate between this model and the standard three-neutrino
oscillation scenario. We 
find that DUNE cannot make a distinction at more than
$3\sigma$ C.L. for bimaximal, tribimaximal and hexagonal mixing, while JUNO can rule out the DM scenario at more than $6\sigma$ C.L. by making high-precision measurements of both $\theta_{12}$ and $\theta_{13}$.

\vskip 0.1in
{\bf Acknowledgements.}
This research was supported in part by the U.S. DOE under
Grant No. DE-SC0010504.

\appendix

\section{Second-order corrections}
\label{ap:second-order}
For $\theta_{23}^0=45^\circ$ and $\theta_{13}^0=0$,
the mass matrix can
be rewritten as
\begin{align}
M &= m_1\text{I} + R_{23}^0\begin{pmatrix}
\delta m_{21}(s_{12}^{0})^2 & \delta m_{21}c_{12}^0s_{12}^0 & \sqrt{2}\epsilon \\
\delta m_{21}c_{12}^0s_{12}^0 & \delta m_{21}(c_{12}^{0})^2 & \epsilon' \\
\sqrt{2}\epsilon & \epsilon' & \delta m_{31}
\end{pmatrix}(R_{23}^0)^T\,.
\label{eq:squarebracket}
\end{align}
We diagonalize the above mass matrix by the unitary matrix,
\begin{align}
U=R_{23}^0U_\delta R_{12}^0R_{12}'\,,
\end{align}
where $R_{ij}^0$ is the rotation matrix in the $i-j$ plane with a
rotation angle $\theta_{ij}^0$, $U_\delta$ is
\begin{equation}
U_\delta=\begin{pmatrix}
1 & 0 & \delta_{13} \\
0 & 1 & \delta_{23} \\
-\delta_{13}^{*} & -\delta_{23}^{*} & 1\\
\end{pmatrix}\,,
\end{equation}
and $R_{12}'$ is
\begin{equation}
R_{12}'=\begin{pmatrix}
1 & \delta_{12} &0 \\
-\delta_{12}^{*} & 1 & 0 \\
0 & 0 & 1\\
\end{pmatrix}\,.
\end{equation}
From Eq.~(\ref{eq:squarebracket}), the leading order corrections in the 1-3 and 2-3 sector are
\begin{align}
\delta_{13}\approx\frac{\sqrt{2}\epsilon^*}{\delta m_{31}}\,,\quad \delta_{23}\approx\frac{\epsilon'^*}{\delta m_{31}}\,.
\label{eq:delta1323}
\end{align}
We see that after the rotations of $R_{23}^0$, $U_\delta$ and $R_{12}^0$, the mass matrix in the 1-2 sector is
\begin{align}
\begin{pmatrix}
0 & 0 \\
0 & \delta m_{21}
\end{pmatrix}
-\frac{1}{\delta m_{31}}\begin{pmatrix}
2\epsilon^2 (c_{12}^{0})^2 +\epsilon'^2(s_{12}^{0})^2-\sqrt{2}\epsilon\epsilon'\sin(2\theta_{12}^0) & \sqrt{2}\epsilon\epsilon'\cos(2\theta_{12}^0)+\frac{2\epsilon^2-\epsilon'^2}{2}\sin(2\theta_{12}^0) \\
\sqrt{2}\epsilon\epsilon'\cos(2\theta_{12}^0)+\frac{2\epsilon^2-\epsilon'^2}{2}\sin(2\theta_{12}^0) & 2\epsilon^2 (s_{12}^{0})^2 +\epsilon'^2(c_{12}^{0})^2-\sqrt{2}\epsilon \epsilon'\sin(2\theta_{12}^0) 
\end{pmatrix}\,.
\end{align}
Hence, the next-to-leading order correction in the 1-2 sector is
\begin{align}
\delta_{12}
&\approx -\frac{\sqrt{2}\epsilon^{*}\epsilon'^{*}\cos(2\theta_{12}^0)+(\epsilon^{*2}-\epsilon'^{*2}/2)\sin(2\theta_{12}^0)}{\delta m_{21}\delta m_{31}}\,,
\end{align} 

Since the DM perturbation potentially introduces additional complex
phases in $\delta\theta_{23}$ and $\delta\theta_{12}$, we must recast the
parameters to put them in the standard form. We combine the
initial rotation with the infinitesimal one to get (e.g., in the 2-3
sector),
\begin{align}
\begin{pmatrix}
c_{23}^0 & s_{23}^0\\ -s_{23}^0 & c_{23}^0
\end{pmatrix}
\begin{pmatrix}
1 & \delta_{23}\\ -\delta_{23}^* & 1
\end{pmatrix} &=
\begin{pmatrix}
c_{23}^0-s_{23}^0\delta_{23}^* & s_{23}^0+c_{23}^0\delta_{23}\\
-s_{23}^0-c_{23}^0\delta_{23}^* & c_{23}^0-s_{23}^0\delta_{23}
\end{pmatrix}\,.
\end{align}
To first order, the magnitude of the 1-1 element is
\begin{align}
|(R_{23})_{11}|
&= \sqrt{(c_{23}^0)^2-2s_{23}^0c_{23}^0{\rm Re}\delta_{23}
	+(s_{23}^0)^2|\delta_{23}|^2}
\nonumber\\
&\simeq c_{23}^0-s_{23}^0{\rm Re}\delta_{23}
\simeq \cos(\theta_{23}^0+{\rm Re}\delta_{23})\,,
\end{align}
and the phase is
\be
{\rm arg}((R_{23})_{11}) \equiv \phi_{23}
\simeq \tan\theta_{23}^0{\rm Im}\delta_{23}\,.
\ee
Likewise for the off-diagonal element,
\begin{align}
|(R_{23})_{12}|
&= \sqrt{(s_{23}^0)^2+2s_{23}^0c_{23}^0{\rm Re}(\delta_{23})
	+(c_{23}^0)^2|\delta_{23}|^2}
\nonumber\\
&\simeq s_{23}^0+c_{23}^0{\rm Re}(\delta_{23})
\simeq \sin(\theta_{23}^0+{\rm Re}(\delta_{23}))\,,
\end{align}
and the phase is
\be
{\rm arg}((R_{23})_{12}) \equiv \phi_{23}'
\simeq \cot\theta_{23}^0{\rm Im}\delta_{23}\,.
\ee
Therefore the rotation in the 2-3 sector is now
\be
R_{23} =
\begin{pmatrix}
	c_{23} e^{i\phi_{23}} & s_{23} e^{i\phi_{23}'}\\
	-s_{23} e^{-i\phi_{23}'} & c_{23} e^{-i\phi_{23}}
\end{pmatrix}\,,
\ee
where $\theta_{23}$ includes a shift in Re($\delta_{23}$). Since
$|\delta\theta_{23}| \ll \theta_{23} \sim 1$, $\phi_{23}$ and
$\phi_{23}^\prime$ are small. A similar manipulation can be done for
$R_{12}$, with $|\phi_{12}|, |\phi_{12}^\prime| \ll 1$. In the 1-3
sector, we get
\begin{align}
R_{13} =
\begin{pmatrix}
1 & \delta_{13}\\
-\delta_{23}^* & 1
\end{pmatrix}
\simeq
\begin{pmatrix}
c_{13} & s_{13}e^{i\phi_{13}}\\
-s_{13}e^{-i\phi_{13}} & c_{13}
\end{pmatrix}\,,
\end{align}
where $\phi_{13} = {\rm arg}(\delta_{13})$. Note that the cosine terms in
$R_{13}$ do not get a phase at first order because their shifts are at second
order (due to the fact that $\theta_{13}^0 = 0$). Hence, the leading corrections to the three mixing angles are
\begin{align}
\delta\theta_{13}&\approx\frac{\sqrt{2}|\epsilon|}{\delta m_{31}}\,,
\\
\delta\theta_{23}&\approx\frac{{\rm Re}(\epsilon^\prime)}{\delta m_{31}}\,,
\\
\delta\theta_{12}&\approx
-\frac{{\rm Re}(\sqrt{2}\epsilon\epsilon^\prime\cos2\theta_{12}^0
	+(\epsilon^2-\epsilon^{\prime2}/2)\sin2\theta_{12}^0)}
{\delta m_{21}\delta m_{31}}\,.
\end{align}

Combining these 2-D rotations together in the full 3-D rotation matrix
and making some phase changes in rows and columns so that the 1-1,
1-2, 2-3, and 3-3 elements are real we get
\begin{align}
U =
\begin{pmatrix}
c_{13}c_{12} & c_{13}s_{12} & s_{13}e^{i\phi_{13}}\\
-s_{12}c_{23}e^{i\eta}-c_{12}s_{23}s_{13}e^{-i\phi_{13}}
& c_{12}c_{23}e^{i\eta}-s_{12}s_{23}s_{13}e^{-i\phi_{13}}
& c_{13}s_{23}\\
s_{12}s_{23}e^{i\eta}-c_{12}c_{23}s_{13}e^{-i\phi_{13}}
& -c_{12}s_{23}e^{i\eta}-s_{12}c_{23}s_{13}e^{-i\phi_{13}}
& c_{13}c_{23}
\end{pmatrix}\,.
\end{align}
where $\eta = \phi_{23}-\phi_{23}^{'}-\phi_{12}-\phi_{12}^{'}$.
This is not quite in the standard form, but we can multiply the second and
third rows by $e^{-i\eta}$ and the third column by $e^{i\eta}$ to get the
standard form for $U$ with $\delta_{CP} = -(\phi_{13}+\eta)$. Since
the phases in $\eta$ are all small, $\delta_{CP}$ is primarily given by
$-\phi_{13}$, i.e., arg($\epsilon$).

\section{A general treatment of oscillation probabilities}
\label{ap:general}

A general way to look at the oscillation probabilities is to do a Taylor series expansion
about the standard expression:
\bea
P &\approx& P_0
+ \left(\frac{\partial P}{\partial\theta_{12}}\right)_0 \delta\theta_{12} C^2
+ \left(\frac{\partial P}{\partial\theta_{13}}\right)_0 \delta\theta_{13} C
+ \left(\frac{\partial P}{\partial\theta_{23}}\right)_0 \delta\theta_{23} C
\\
&\phantom{\approx}&
+ \frac{1}{2}\left[
\left(\frac{\partial^2 P}{\partial\theta_{12}^2}\right)_0 \delta\theta_{12}^2 C^4
+ \left(\frac{\partial^2 P}{\partial\theta_{13}^2}\right)_0 \delta\theta_{13}^2 C^2
+ \left(\frac{\partial^2 P}{\partial\theta_{23}^2}\right)_0 \delta\theta_{23}^2 C^2\right]
\nonumber\\
&\phantom{\approx}&
+ \left(\frac{\partial^2 P}{\partial\theta_{12}\partial\theta_{13}}\right)_0 \delta\theta_{12}\delta\theta_{13} C^3
+ \left(\frac{\partial^2 P}{\partial\theta_{12}\partial\theta_{23}}\right)_0 \delta\theta_{12}\delta\theta_{23} C^3
+ \left(\frac{\partial^2 P}{\partial\theta_{13}\partial\theta_{23}}\right)_0 \delta\theta_{13}\delta\theta_{23} C^2\,.
\nonumber
\label{eq:taylorexpansion}
\eea
Using $\langle C \rangle = 0$,
$\langle C^2 \rangle = 1/2$, $\langle C^3 \rangle = 0$, and $\langle
C^4 \rangle = 3/8$, where $\langle~\rangle$ indicates averaging over
the DM oscillation, and after averaging,
\bea
\label{eq:afterave}
\langle P \rangle &\approx& P_0
+ \frac{1}{2}\left(\frac{\partial P}{\partial\theta_{12}}\right)_0 \delta\theta_{12}
+ \frac{3}{16} \left(\frac{\partial^2 P}{\partial\theta_{12}^2}\right)_0 \delta\theta_{12}^2
\\
&\phantom{\approx}&
+ \frac{1}{4}\left(\frac{\partial^2 P}{\partial\theta_{13}^2}\right)_0 \delta\theta_{13}^2
+ \frac{1}{4}\left(\frac{\partial^2 P}{\partial\theta_{23}^2}\right)_0 \delta\theta_{23}^2
+ \frac{1}{2}\left(\frac{\partial^2 P}{\partial\theta_{13}\partial\theta_{23}}\right)_0 \delta\theta_{13}\delta\theta_{23}\,.
\nonumber
\eea
On the other hand, the expansion in terms of effective parameter shifts is
\bea
\label{eq:eff}
P &\approx& P_0
+ \left(\frac{\partial P}{\partial\theta_{12}}\right)_0 \delta\theta_{12}^{eff}
+ \left(\frac{\partial P}{\partial\theta_{13}}\right)_0 \delta\theta_{13}^{eff}
+ \left(\frac{\partial P}{\partial\theta_{23}}\right)_0 \delta\theta_{23}^{eff}
\\
&\phantom{\approx}&
+ \frac{1}{2}\left[
\left(\frac{\partial^2 P}{\partial\theta_{12}^2}\right)_0 (\delta\theta_{12}^{eff})^2
+ \left(\frac{\partial^2 P}{\partial\theta_{13}^2}\right)_0 (\delta\theta_{13}^{eff})^2
+ \left(\frac{\partial^2 P}{\partial\theta_{23}^2}\right)_0 (\delta\theta_{23}^{eff})^2\right]
\nonumber\\
&\phantom{\approx}&
+ \left(\frac{\partial^2 P}{\partial\theta_{12}\partial\theta_{13}}\right)_0 \delta\theta_{12}^{eff}\delta\theta_{13}^{eff}
+ \left(\frac{\partial^2 P}{\partial\theta_{12}\partial\theta_{23}}\right)_0 \delta\theta_{12}^{eff}\delta\theta_{23}^{eff}
+ \left(\frac{\partial^2 P}{\partial\theta_{13}\partial\theta_{23}}\right)_0 \delta\theta_{13}^{eff}\delta\theta_{23}^{eff}\,.
\nonumber
\eea

If the expressions are quadratic in $s_{13}^2$ and/or $c_{13}^2$ and using
$\theta_{13}^0 = 0$, then $(\partial P/\partial\theta_{13})_0 = 0$,
$(\partial^2 P/\partial\theta_{13}\partial\theta_{23})_0 = 0$, and Eq.~(\ref{eq:afterave})
reduces to (keeping only the leading correction for each $\delta\theta$)
\bea
\label{eq:redave}
\langle P \rangle &\approx& P_0
+ \frac{1}{2}\left(\frac{\partial P}{\partial\theta_{12}}\right)_0 \delta\theta_{12}
+ \frac{1}{4}\left(\frac{\partial^2 P}{\partial\theta_{13}^2}\right)_0 \delta\theta_{13}^2
+ \frac{1}{4}\left(\frac{\partial^2 P}{\partial\theta_{23}^2}\right)_0 \delta\theta_{23}^2\,,
\eea
and Eq.~(\ref{eq:eff}) reduces to (again keeping only the leading correction for each
$\delta\theta)$
\bea
\label{eq:redeff}
P &\approx& P_0
+ \left(\frac{\partial P}{\partial\theta_{12}}\right)_0 \delta\theta_{12}^{eff}
+ \left(\frac{\partial P}{\partial\theta_{23}}\right)_0 \delta\theta_{23}^{eff}
+ \frac{1}{2}\left(\frac{\partial^2 P}{\partial\theta_{13}^2}\right)_0 (\delta\theta_{13}^{eff})^2\,.
\eea
Comparing Eqs.~(\ref{eq:redave}) and~(\ref{eq:redeff}), we find $\delta\theta_{12}^{eff} =
\delta\theta_{12}/2$, $\delta\theta_{13}^{eff} = \delta\theta_{13}/\sqrt2$,
and $\delta\theta_{23}^{eff} = 0$ (neglecting the small, second-order correction
to $\theta_{23}$, which is acceptable since the leading order terms
involving $\theta_{23}$ are generally not zero). Note that since the period of DM oscillation considered here is much larger than the neutrino travel time at a terrestrial experiment, the expansions in Eqs.~(\ref{eq:taylorexpansion}) and (\ref{eq:eff}) are not affected by matter effects and the shifts in the effective angles remain the same.

In the more general case with $(\partial P/\partial\theta_{13})_0 \ne
0$ (such as when there is a single factor of $s_{13}$ or
$\sin2\theta_{13}$), there is no single power of
$\delta\theta_{13}$ in Eq.~(\ref{eq:afterave}) that matches the single power of
$\delta\theta_{13}^{eff}$ in Eq.~(\ref{eq:eff}), and the simple correspondence
between $\delta\theta_{13}$ and $\delta\theta_{13}^{eff}$ breaks down.
The only measurement that appears to have this problem is the appearance measurement at long-baseline experiments.
Also, since the Dirac CP phase is always associated with $s_{13}$ in an oscillation probability, the absence of a single power of $\delta\theta_{13}$ in Eq.~(\ref{eq:afterave}) indicates a reduced sensitivity to the Dirac CP phase in neutrino oscillation experiments.


\end{document}